\documentclass{article}

\usepackage{geometry}
\usepackage{blindtext}
\usepackage[absolute]{textpos}
\usepackage[super]{natbib}
\usepackage{url}
\usepackage{soul}
\usepackage{longtable,booktabs}
\usepackage{afterpage}
\usepackage{amsmath,amssymb,amsfonts}
\usepackage{algorithm}
\usepackage{algorithmic}
\usepackage{array, makecell} %
\usepackage{graphicx}
\usepackage{gensymb}
\usepackage{textcomp}
\usepackage{multirow}
\usepackage{xcolor}
\usepackage[para,online,flushleft]{threeparttable}
\graphicspath{{images/},{.}}
\usepackage{supertabular}
\usepackage{url}
\def\UrlBreaks{\do\/\do-}
\usepackage[
  separate-uncertainty = true,
  multi-part-units = repeat
]{siunitx}
\DeclareGraphicsExtensions{.pdf,.jpeg,.png,.jpg}
\def\BibTeX{{\rm B\kern-.05em{\sc i\kern-.025em b}\kern-.08em
    T\kern-.1667em\lower.7ex\hbox{E}\kern-.125emX}}
\usepackage{authblk}

\title{Bird-Area Water-Bodies Dataset (BAWD) and Predictive AI Model for Avian Botulism Outbreak (AVI-BoT)}
\begin{document}

\begin{textblock}{20}(1,1)
\noindent\Large A version of this manuscript is under review and consideration for publication
\end{textblock}



\author{Narayani Bhatia, Devang Mahesh, Jashandeep Singh, and Manan Suri*}

\affil{Department of Electrical Engineering, Indian Institute of Technology Delhi, India. \\ *Corresponding author: Manan Suri (e-mail: manansuri@ee.iitd.ac.in).}
\date{}
\maketitle
\begin{abstract}
Avian botulism is a paralytic bacterial disease in birds often leading to high fatality. In-vitro diagnostic techniques such as Mouse Bioassay, ELISA, PCR are usually non-preventive, post-mortem in nature, and require invasive sample collection from affected sites or dead birds. In this study, we build a first-ever multi-spectral, remote-sensing imagery based global \textit{Bird-Area Water-bodies Dataset (BAWD)}  (i.e. fused satellite images of warm-water lakes/marshy-lands or similar water-body sites that are important for avian fauna) backed by on-ground reporting evidence of outbreaks. BAWD consists of 16 topographically diverse global sites monitored over a time-span of 4 years (2016-2021). We propose a first-ever Artificial Intelligence based (AI) model to predict potential outbreak of Avian botulism called AVI-BoT (\textbf{A}erosol \textbf{V}isible, \textbf{I}nfra-red (NIR/SWIR) and \textbf{B}ands of \textbf{T}hermal). We also train and investigate a simpler (5-band) Causative-Factor model (based on prominent physiological factors reported in literature) to predict Avian botulism. AVI-BoT demonstrates a training accuracy of 0.96 and validation accuracy of 0.989 on BAWD, far superior in comparison to our model based on causative factors. We also perform an ablation study and perform a detailed feature-space analysis. We further analyze three test case study locations - Lower Klamath National Wildlife Refuge and Langvlei and Rondevlei lakes where an outbreak had occurred, and Pong Dam where an outbreak had not occurred and confirm predictions with on-ground reportings. The proposed technique presents a scale-able, low-cost, non-invasive methodology for continuous monitoring of bird-habitats against botulism outbreaks with the potential of saving valuable fauna lives.

\end{abstract}



\maketitle


\section{Introduction}

\begin{figure}[htb]
\centering
\includegraphics[width=1.1\linewidth]{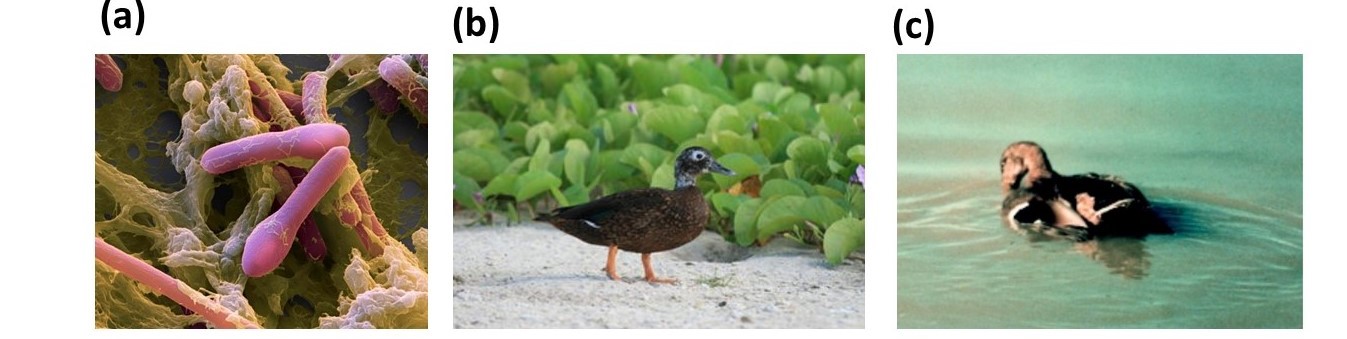}
\caption{(a) Bacterium - Clostridium botulinum\cite{bacteria}, (b) Healthy bird \cite{healthy_bird}, (c) Affected bird suffering from paralysis of neck muscles due to Avian botulism \cite{death_stats}.}
\label{birds}
\end{figure}

Avian botulism is a paralytic disease that fatally affects birds. It is caused by a bacterium, Clostridium botulinum, which produces a Botulinum neurotoxin, ingestion of which often leads to death in birds (Fig. \ref{birds}). Botulism spores (resting stage of the bacteria) can persist in the soil and aquatic sediments of water bodies for decades before it enters the food webs of birds. Organisms like- algae, plants and invertebrates act as biotic reservoir for the disease and further fishes are known carriers of botulism spores\cite{negative_ref}. Once the spores enter the food webs of birds, avian botulism spreads in a self-perpetuating manner via the maggot cycle. Maggots which feed on the dead birds, acquire the bacteria and are likely to be eaten by carnivorous birds. The carnivorous birds that die because of the bacteria present in maggots are then fed by other maggots and this way, the cycle continues. Avian botulism is prevalent all over the world, with higher prevalence over North American wetlands. In the previous century the disease has caused an estimated average death of about 644,000 birds per reported outbreak year in U.S.A. and Canada {\cite{death_stats}} alone. Some physiological conditions for the bacterium to develop are: low-oxygen, eutrophic zones, higher water temperatures, shallow stagnant water {\cite{holmes2019avian}}. Such non-exhaustive list of causative factors leading to Avian botulism outbreak is one of the key areas of current research \cite{anza2014eutrophication}. Some of the recent outbreaks in England (2018), Wales (2018) \cite{holmes2019avian}, Lower Klamath National Wildlife Refuge (2020) \cite{klamath_news} resulted in bird casualty as high as 40,000.

Some of the widely known techniques to confirm the outbreak of Avian botulism are culture techniques \cite{culture}, assay test for presence of toxin \cite{rocke1998preliminary}, PCR test \cite{chellapandi2018pcr} (See Table \ref{table_lit_review}). Most of these techniques rely on actual ground level sample collection which is challenging due to the requirement of physical access. Further, all the aforementioned techniques are usually used as post-mortem tests \cite{anniballi2013management}, that is, used for detection of conditions only after clinical symptoms or deaths start getting reported from the location. Moreover, in some cases, post-mortem tests may be inconclusive and need to be backed by further laboratory confirmation which adds to the delay \cite{anniballi2013management}. This often leads governments and conservationists with little time for quick relief operations to save precious and endangered fauna. 

In order to overcome some limitations of the clinical diagnostic methods, we present in this study a first-ever \textit{Bird-Area Water-Bodies Dataset (BAWD)} (i.e. fused satellite images of warm-water lakes/marshy-land etc or similar water-body sites that are important for avian fauna) built from publicly available open-source satellite imagery. Our hand-annotated dataset consists of multi-spectral satellite images, covering a total ground area of 1039 sq. km, from two open source satellite projects (Sentinel and Landsat), covering 16 topographically diverse global bird-area water bodies from 4 continents, where locations are monitored over a time-span of 4 years (2016-2021). BAWD includes complete mapping between location scenes and ground-truth labels, backed by reporting of outbreaks.

Further, we propose a multi-spectral earth observation supervised CNN based detection model called AVI-BoT (\textbf{A}erosol \textbf{V}isible \textbf{I}nfra-red (both NIR and SWIR) - \textbf{B}ands of \textbf{T}hermal) to predict potential outbreak of Avian botulism. AVI-BoT takes fused multi-spectral satellite images as an input and generates a spatial prediction map depicting probability of an outbreak to occur.
Satellite images in BAWD are composed of 10 remote sensing bands which capture the physio-chemical botulinum-promoting parameters such as pH, salinity, temperature among many others. We also train and investigate a simpler (5-band) Causative-Factor model (based on prominent physiological factors reported in literature as conducive for outbreak) to predict Avian botulism outbreak.

\begin{table}[htbp]
\centering
\renewcommand{\arraystretch}{1.1}
\caption{State of the Art techniques for detection of Avian botulism.}
\label{table_lit_review}
\makebox[\linewidth]{
\begin{tabular}{|p{1.5cm}|p{2cm}|p{3cm}| p{4cm} | p{4cm}|} 
\hline
Paper & Objective of Study & Locations Covered & Underlying Methodology & Key Contribution \\
\hline
\cite{mousebioassay} & Detection & 2 locations & Mouse BioAssay (Clinical, Bio-chemical) & Identification of the neurotoxin produced by Clostridium botulinum using mouse bioassay technique\\
\hline
\cite{rocke1998preliminary, ELISA_AB} & Detection & - & ELISA (Clinical, Bio-chemical) & Identification of the neurotoxin produced by Clostridium botulinum using microplates\\
\hline
\cite{chellapandi2018pcr} & Detection & - & PCR (Clinical, Bio-chemical) & Reviews PCR-based assay along with primers, sensitivity\\
\hline
\cite{le2017development} & Detection & - & Pre-PCR (Clinical, Bio-chemical) & Optimized pre-PCR processing by proposing the use of liver\\
\hline
\textbf{This work} & \textbf{Detection, Prediction of potential outbreaks} & \textbf{16 locations}, \textbf{4 continents} & \textbf{Salinity, Temperature, pH, Organic Matter (Earth observation based Computational-Deep Learning Network)} & \textbf{Proposed use of AVI-BoT Model using temporal multispectral remote sensing data of 10-100m resolution from Sentinel 2 and Landsat 8}\\
\hline
\end{tabular}}
\end{table}

Further, three test locations at Lower Klamath National Wildlife Refuge (U.S.A., October 2020), Langvlei and Rondevlei Lakes (South Africa, January, February, May 2017) and Pong Dam (India, January 2021) are investigated using AVI-BoT. At the two former positive locations, AVI-BoT provides accurate forecasting of the outbreak, validated using field reporting and correctly predicts the latter location to be negative. We then carry out temporal analysis for Lower Klamath National Wildlife Refuge to investigate the formation of Avian botulism conducive conditions which exacerbated and resulted in mass die-offs. We also validate the predictions on Langvlei and Rondevlei Lakes with the on-ground actual reporting of bird mortality from a study conducted in 2017. 

To take a closer look at the evolution of the feature representations learnt by AVI-BoT, we additionally train four direct spectral models: ((i) Aerosol, (ii) Visible (3-bands), (iii) IR (3-bands), and (iv) Thermal (3-bands)).

\section{Methods}\label{methods}
\subsection{\textit{Bird-Area Water-Bodies} Dataset (BAWD)}\label{dataset_bawd}
\begin{figure}[H]
\centering
\includegraphics[width=1.1\linewidth]{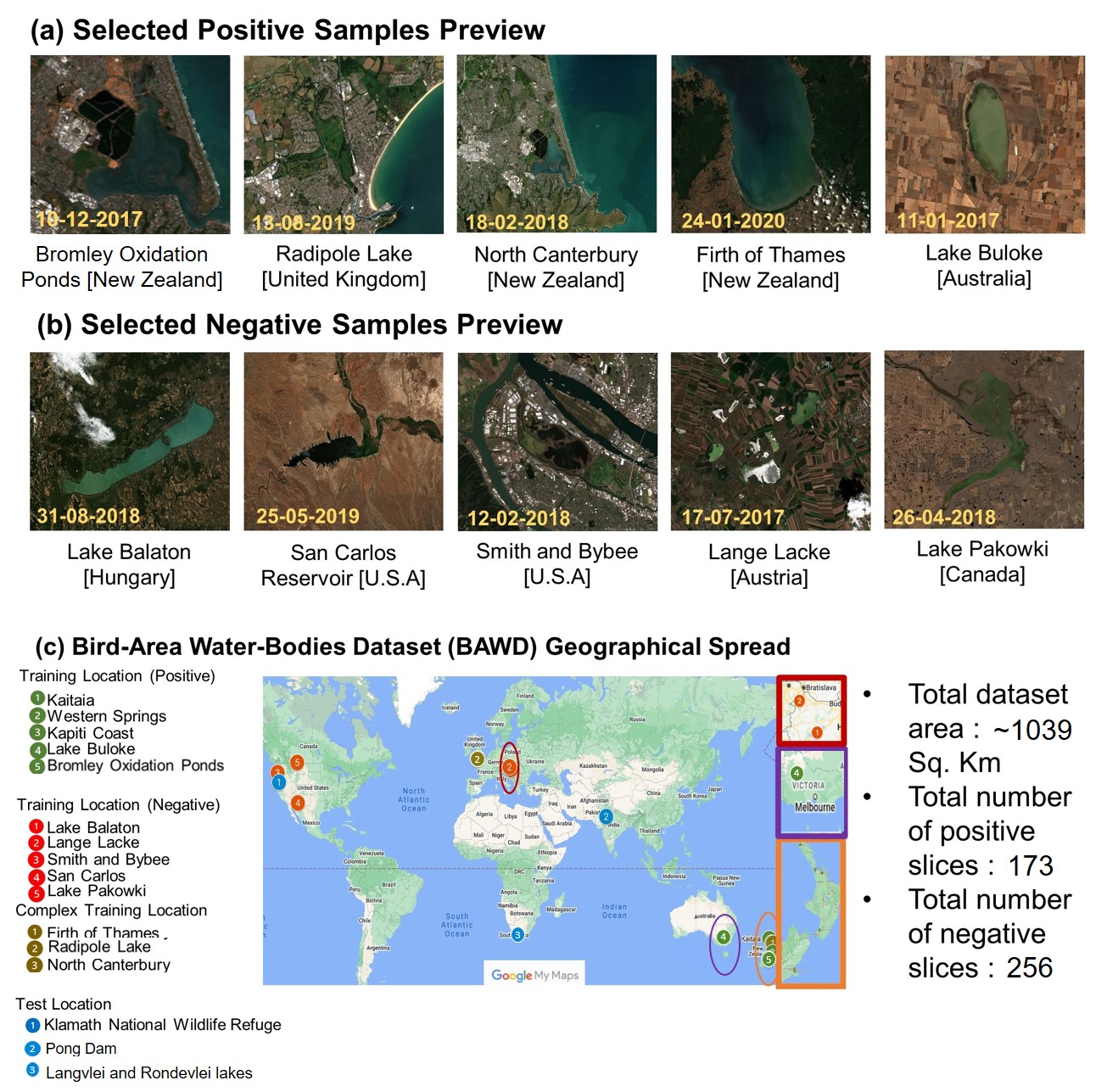}
\caption{\textit{Bird-Area Water-Bodies Dataset (BAWD)} details: RGB preview of few (a) Positive and (b) Negative samples (Raw data source: Modified Copernicus Sentinel data 2016-2020 /Sentinel Hub), (c) Distribution map of 16 global locations chosen for the study. }
\label{dataset_pn}
\end{figure}

A multi-spectral spatio-temporal dataset spanning across 16 topographically diverse global locations (13 training and 3 test) comprising a total of 446 satellite images (429 training and 17 test images) of water bodies (rivers, lakes, ponds and wetlands) which are bird habitats, was built and manually annotated for this study (see Fig. \ref{dataset_pn}). Fig. \ref{dataset_pn} (c) shows the spatial variability of the dataset. The images have been generated from Sentinel EO browser\cite{Sentinel_EO_browser}. Among the 10 bands, 7 bands are raw spectral bands taken from Sentinel 2A after L2A processing. L2A processing takes as input L1C processed image which is radio-metrically and geometrically corrected (including orthorectification and spatial registration) and performs atmospheric corrections \cite{l2aproc}. The remaining 3 bands are thermal bands which have been taken from Landsat 8 or Sentinel 3. Sentinel-3 data was used wherever Landsat-8 data was cloudy or not available for dates of interest. Landsat 8 is preferred for thermal maps owing to its superior spatial resolution over Sentinel 3. 

For building a robust model capable of predicting Avian botulism outbreak near/in water-bodies across seasonal variability, we collected positive and negative data spanning over a time-period of $\sim$ 3 years. Positive-labels denote conditions where Avian botulism exist while negative-labels denote non-existent botulism conditions. Positive labels were created on the basis of confirmed ground reports for the outbreak at any of the sites. The temporal spread of each positive label was adjusted over a span of three months (one month before and after) with respect to the ground reporting date of the outbreak since the conditions do not change overnight. 

For samples labelled as negative, the locations are predominantly chosen from the list of Important Bird Areas (IBA) designated by BirdLife International \cite{birdlife}. In some cases, confirmed news reportings stating a corrective action taken by the local authorities are also used for negative samples. Corrective actions include draining or flooding the wetland to change the environmental conditions sufficiently to stop the production of toxin by Clostridium botulinum. Three complex locations which have transitioned between 'outbreak' and 'no outbreak' have also been factored in the BAWD training dataset to account for complex cases. Table \ref{table_dataset_p} details the locations and the dates for which the samples are collected along with ground reporting evidence for labelling the sample-type (i.e. positive or negative).

\begin{table}
\caption{Description of Positive and Negative Samples in \textit{Bird-Area Water Bodies} (BAWD) Training Dataset (13 out of 16 locations).}
\label{table_dataset_p}
\centering
\makebox[\linewidth]{
\begin{tabular}{|p{4cm}|c|p{2cm}|p{2cm}|}
\hline
\multicolumn{4}{c}{\textbf{Positive Data}}\\
\hline
Location & Period of data collection & Number of images & Reported positive in: \\
\hline
Bromley Oxidation Pond (New Zealand) & Dec'17 - Mar'18 & 5 & \cite{bromley_news} \\
\hline
Firth of Thames (New Zealand) & Dec'19 - Feb'20 & 99 & \cite{firth_news} \\
\hline
Kaitaia (New Zealand) & Jan'19 - Mar'19 & 7 &\cite{kaitaia_news}\\ 
\hline
Western Springs (New Zealand) & Jan'18 - Feb'18 & 1 & \cite{western_news} \\ 
\hline
Kapiti Coast (New Zealand) & Dec'17 - Mar'18 & 7 &\cite{kapiti_news} \\ 
\hline
Lake Buloke (Australia) & Jan'17 - Mar'17 & 6 &\cite{buloke_news} \\
\hline
North Canterbury (New Zealand) & Jan'18 - Apr'18 & 18 & \cite{north_news}\\
\hline
Radipole Lake (United Kingdom)& Aug'19 - Sept'19 & 30 & \cite{radipole_news}\\
\hline
\multicolumn{4}{c}{\textbf{Negative Data}}\\
\hline
Location & Period of data collection & Number of images & Reported negative in: \\
\hline
Firth of Thames (New Zealand) & June'17 ; Aug'18 ; June'19 - Aug'19 & 51 & \cite{firth_iba} \\
\hline
Lake Balaton (Hungary) & Aug'18 - Sept'18 ; July'19 - Sept'19 & 62 & \cite{negative_ref, balaton_iba} \\
\hline
Lange Lacke (Austria) & Dec'16 - Feb'17 ; July'17-Sept'17 & 24 &\cite{lange_iba} \\
\hline
Smith and Bybee (U.S.A.) & Feb'18 -Mar'18 ;Feb'19 - Apr'19 & 26 & \cite{smith_bybee_proposal} \\
\hline 
San Carlos Reservoir (U.S.A.) & Jan'19 - Dec'19 & 17 &\cite{sanc_news}  \\
\hline
Radipole Lake (United Kingdom) & Jan'18 - May'18; July'18 - Dec'18 & 12 & \cite{radipole_iba}\\
\hline
North Canterbury (New Zealand) & Jan'17-Feb'17; May'17; July'17; Oct'17-Dec'17 & 10 & \cite{nc_iba} \\
\hline
Lake Pakowki (Canada) & Apr'18-Aug'18; Oct'18-Nov'18; Apr'19-Aug'19 & 54 & \cite{pak_iba} \\
\hline

\end{tabular}}
\end{table}

\subsection{CNN Training Methodology}

Each of the 10 bands (Table \ref{table_dataset_bands}) is merged using QGIS\cite{qgis}, and water area is extracted for each test case manually. The merge operation is carried out such that, the lowest resolution among the Sentinel 2A constituent bands is assigned to the output merged file. As shown in Table \ref{table_dataset_bands}, Aerosol band (B01) has the lowest resolution among the Sentinel 2A bands, that is 60 m and it is assigned to the artificially synthesised data point. Further, water extraction can also be carried out by an automated deep-learning based segmentation network like UNet \cite{unet}. Selective bands from this artificial data point serve as the input to the CNN in the various experiments.

\begin{table}
\caption{Details of spectral-bands chosen and the represented parameters to generate proposed \textit{Bird-Area Water Bodies} dataset.}
\label{table_dataset_bands}
\centering
\makebox[\linewidth]{
\begin{tabular}{|c|c|c|p{2cm}|p{2cm}|p{5cm}|p{2cm}|}
\hline
Sensor & Band & Description & Wavelength ($\mu$m) & Resolution (m)  & Parameters  & Dependence\\
\hline
S2 & 1 & Coastal/Aerosol & 0.443 & 60 & pH level, Redox potential, Presence of microbes \& Water constituents & \cite{karaoui2019evaluating, pizani2020estimation}\\
\hline
S2 & 2 & Blue & 0.490 & 10 & Chlorophyll concentration level, DOM, Salinity level & \cite{pizani2020estimation, dom, salinity}\\
\hline
S2 & 3 & Green & 0.560 & 10 & pH level, Chlorophyll concentration level, DOM, Salinity level &   \cite{karaoui2019evaluating, pizani2020estimation, dom, salinity}\\
\hline
S2 & 4 & Red & 0.665 & 10 &  pH level, Chlorophyll concentration level, Redox potential & \cite{karaoui2019evaluating, pizani2020estimation, torres2020water}\\
\hline
S2 & 8a & \makecell{Vegetation \\ Red Edge} & 0.865 & 20 & pH level, Vegetation \& Organic Matter & \cite{torres2020water} \\
\hline
S2 & 11 & SWIR & 1.610 & 20 & Dissolved Oxygen level, Vegetation \& Organic Matter & \cite{pizani2020estimation, karaoui2019evaluating}\\
\hline
S2 & 12 & SWIR & 2.190 & 20 & Vegetation \& Organic Matter & \\
\hline
L8/S3 & R & Temperature$^\star$ & - & 100/1k & Surface temperature & \cite{level_temp}\\
\hline
L8/S3 & G & Temperature$^\star$ & - & 100/1k & Surface temperature & \cite{level_temp}\\
\hline
L8/S3 & B & Temperature$^\star$ & - & 100/1k & Surface temperature & \cite{level_temp}\\
\hline
\end{tabular}}
\begin{tablenotes}
      \small
      \item [$\star$] Causative Factors based model consists of 5 bands- algebraically calculated: (i) Salinity (ii) DOM and raw spectral bands (iii)-(v) Thermal bands \\
\end{tablenotes}

\end{table}

All models trained in this study use a binary classification deep CNN with AlexNet \cite{krizhevsky2017imagenet} as the base architecture. It comprises of 5 convolutional layers followed by 2 fully connected layers. For the purpose of training, each image in the dataset is divided into 31x31 chips. We get 942 chips from positive images and 2064 chips from negative images in the dataset. To account for the mismatch in number of positive and negative samples, higher weight is given to positive samples than negative samples in the loss function (shown in Eq.\ref{loss_fn}).
\begin{equation}\label{loss_fn}
\begin{split}
    W.B.C.E = - [ & w_1*y_{true}.log(y_{pred}) + \\  
    & w_0*(1-y_{true})*log(1 - y_{pred}) ]  \\ 
\end{split}
\end{equation}

Where:\\ 
    $W.B.C.E$: Weighted Binary Cross Entropy Loss per pixel\\
    $y_{true}$: True label for a pixel \\
    $y_{pred}$: Predicted label for a pixel\\ 
    $w_{1}$: Weight assigned to positive training data\\ 
    $w_{0}$: Weight assigned to negative training data\\ 
    Since the proportion of chips labelled as positive:negative is $942:2064 = \sim 3:7 $, thus the mismatch is accounted for by assigning $w_{1}=7$ and $w_{0}=3$.
 Further, the chips obtained are divided into $\sim$ 10:1 train : validation sets. The training parameters chosen are shown in Fig. \ref{fig_flowchart}.

\begin{algorithm}[h!]
\begin{algorithmic} 
\REQUIRE{Current Epoch $e_i$, Base Epoch $E_{base}$, Step Epoch $E_{step}$, Max Epoch $E_{max}$, Base Learning Rate $LR_{base}$}
\ENSURE{New Learning Rate $LR_{new}$}
\WHILE{$e_i < E_{max} $}
\IF{$e_i< E_{base} + E_{step}$}
\STATE{$LR_{new}=LR_{base}$}
\ELSE
\STATE{$LR_{new}=LR_{base} \times 0.1 $}
\STATE{$LR_{base} \gets LR_{new} $}
\STATE{$E_{base} \gets E_{base} + E_{step} $}
\ENDIF
\ENDWHILE
\end{algorithmic}
\caption{Learning rate scheduler algorithm inspired by combination of techniques in \cite{learningrate}.}
\label{algo2}
\end{algorithm}

The output of the network is a probability value that indicates the likelihood for occurrence of Avian botulism outbreak (ranging from 0 to 1). The output probability value for a chip is assigned to the center pixel of the chip. At the time of prediction, we divide an entire image of test location into 31x31 chips with stride size 1. Thus, the entire image is scanned and each pixel is assigned a probability. During the inference, care has been taken to extract the water mask manually using QGIS. As there is no one index that can be used to denote the presence of water especially in complex locations with presence of biomass in water body, we have used Normalized Difference Water Index (NDWI) as a guide to draw the manual mask. The NDWI map is calculated by Sentinel EO browser using the equation shown below:

\begin{equation}\label{ndwi}
\begin{split}
    NDWI =  \frac{Band3 - Band8}{Band3 + Band8}
\end{split}
\end{equation}

where Band3 and Band8 are the respective bands captured by the Sentinel 2 satellite. Band3 corresponds to the Green Band and has a central wavelength of 0.56 $\mu$m and Band8 corresponds to NIR band and has a central wavelength of 0.842 $\mu$m. Sentinel EO browser considers a value of NDWI greater than 0.5 as a water body.

\subsubsection{Causative Factors based Model}\label{causative}
In this model, we combine the most crucial causative factors responsible for the outbreak in literature \cite{10.2307/3802842} i.e. salinity, temperature and dissolved organic matter (DOM). Equivalent band representations for the same are calculated as shown below:

\begin{enumerate}
    \item Salinity\cite{salinity} = $ 10^{0.037X+1.494}$  \\  where $X = ( Band2 -Band3)/(Band2 + Band3)$
    \item DOM\cite{dom} = $0.9819X + 5.6831$ \\ $X = Band 2/Band3$  
    \item Temperature = taken from thermal heat maps
\end{enumerate}

Where Band2 and Band3 refer to Sentinel Bands as defined in Table \ref{table_dataset_bands}.\\

\subsubsection{AVI-BoT Model}
\begin{figure}[H]
\centering
\makebox[1.1\linewidth]{
\includegraphics[height=8in]{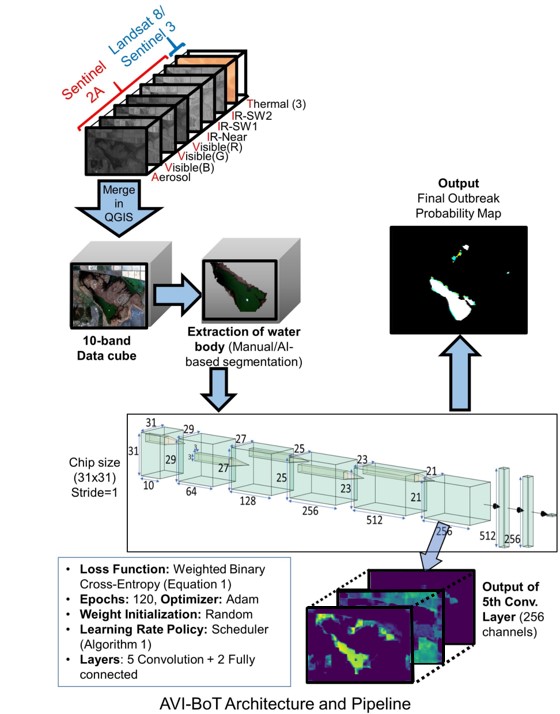}}
\caption{Proposed dataflow and deep-learning backbone used for training and inference for different networks in this study. All 10 bands are used in AVI-BoT.}
\label{fig_flowchart}
\end{figure}

Post exploration of the causative factors based model, we fuse multi-spectral bands from two different satellites to capture the impact of multiple relevant physio-chemical parameters listed in Table \ref{table_dataset_bands}. Hence, we propose a multi-spectral earth observation based supervised CNN detection model called AVI-BoT (\textbf{A}erosol \textbf{V}isible \textbf{I}nfra-red (both NIR and SWIR) - \textbf{B}ands of \textbf{T}hermal) to predict potential outbreak of Avian botulism. While maintaining the tradeoff between the model size and useful features, the 10 constituent bands are carefully chosen to represent each of the wavelength bands (A,V,I,T) to remove redundancy while keeping the model size bounded. The bands - Aerosol (B01), Visible (B02-B04), NIR (B08a)-SWIR (B11,B12), Thermal (Heat maps) - capture physical, chemical and biological parameters where each of these respective `causative factors' would contribute towards the decision of a potential `outbreak' or `no-outbreak'. Fig. \ref{fig_flowchart} depicts the methodology to synthesise the fused input for AVI-BoT Model and to generate a prediction on any test location.  

\section{Results}
\subsection{CNN Model Bench-marking}

\begin{figure}[H]
\centering
\makebox[1.1\linewidth]{
\includegraphics[height=4in]{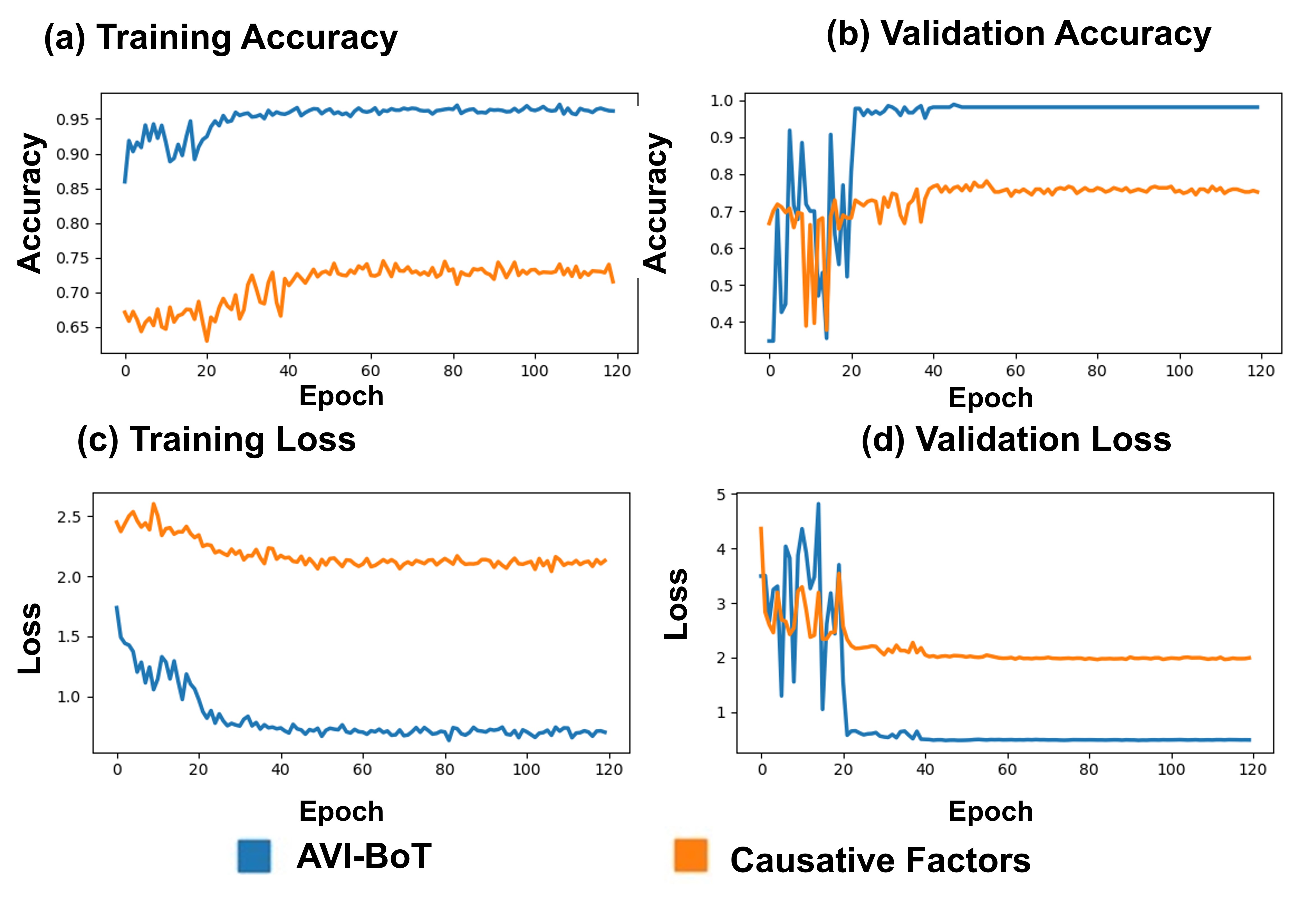}}
\caption{Epoch-wise scores for proposed Causative Factors and AVI-BoT models: (a) Training Accuracy, (b) Validation Accuracy, (c) Training Loss, and (d) Validation Loss.}
\label{fig_train}
\end{figure}

Fig. \ref{fig_train} shows the accuracy and loss plots for training and validation for the two proposed models: (i) Causative Factors based model (5-bands) and (ii) AVI-BoT model (10-bands). The Causative factors model based on factors inspired from literature shows inferior training and validation accuracy compared to AVI-BoT model, thereby necessitating the inclusion of multiple diverse spectral features (detailed in Table \ref{table_dataset_bands}). The 10-band AVI-BoT model (which is a combination of aerosol, visible-spectra, NIR-SWIR, and thermal bands) capably combines all hypothesized factors in the feature representation and achieves a training accuracy of 0.96 and a validation accuracy of 0.989.   

\subsection{Test Case-Studies}
To test the capability of the proposed AVI-BoT model, we analyzed three test locations. These locations include a variety of avian botulism situations - both positive and negative - to perform robust testing of the AVI-BoT model, as two of these locations show high incidence of Avian Botulism and the third was speculated to be a case of Avian Botulism before the final diagnosis came to be Avian Flu. The complexity of these locations arose from the dynamic and evolving on-ground situation, unexpectedly worsening within a matter of days, affecting a large number of birds before authorities could take action. It is noteworthy that in both the positive cases, none of the on-ground apparatus had reported even an inclination towards an impending outbreak. These test case-studies include:

\begin{itemize}
    \item Lower Klamath National Wildlife Refuge, California, USA in October 2020 \cite{klamath_news} - Positive.
    \item Pong Dam, Himachal Pradesh, India in January 2021 \cite{pong_news} - Negative.
    \item Langvlei and Rondevlei lakes, South Africa in January, February and May 2017 \cite{govender2019outbreak} - Positive and then gradually Negative.
\end{itemize}

\subsubsection{Test Case Study I - Lower Klamath National Wildlife Refuge, USA}
\begin{figure}[H]
\makebox[1.1\linewidth]{
\includegraphics[width=7in]{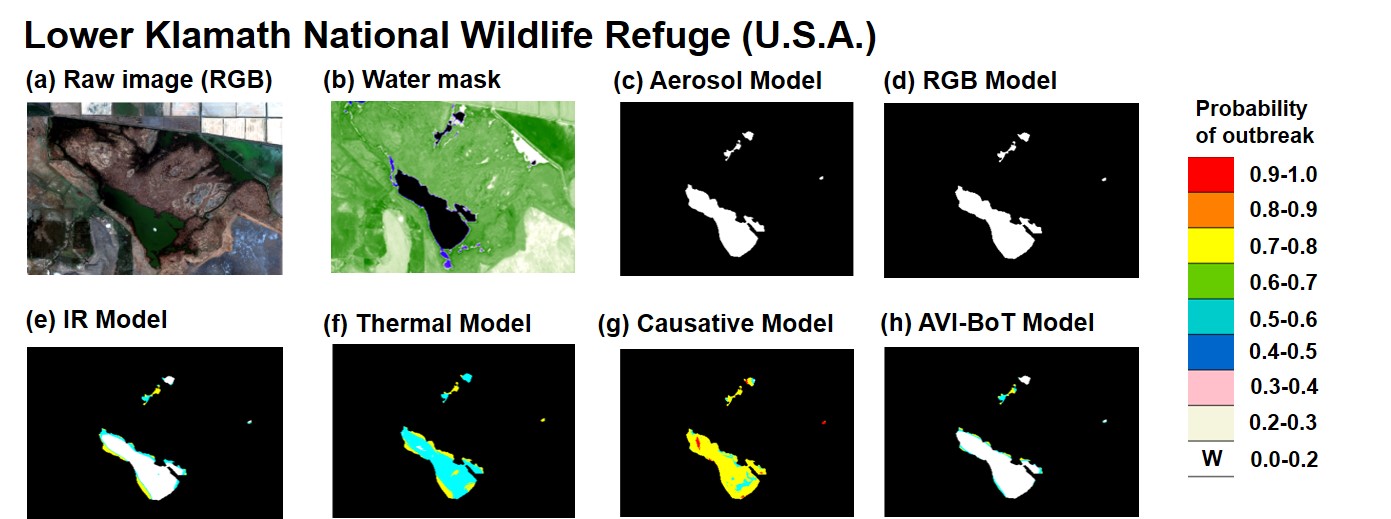}}
\caption{Prediction results on Lower Klamath National Wildlife Refuge, U.S.A. for 9th October 2020 ((a)-(h)).}
\label{fig_all}
\end{figure}

Lower Klamath National Wildlife Refuge (U.S.A.) witnessed an outbreak in October 2020 \cite{klamath_news}. For this case study, we have conducted tests on Sheepy Lake, which forms a portion of the Lower Klamath National Wildlife Refuge. Fig. \ref{fig_all} (b) shows the water mask used for prediction in confirmation with NDWI mask. Aerosol model (Fig. \ref{fig_all} (c)) and Visible model (Fig. \ref{fig_all} (d)) do not yield a positive prediction while Thermal model (Fig. \ref{fig_all} (f)) and Causative Factors based model (Fig. \ref{fig_all} (g)) predict a blanket positive prediction. Only IR Model (Fig. \ref{fig_all} (e)) and AVI-BoT model (Fig. \ref{fig_all} (h)) predict the outbreak with spatial granularity. Another important point to note in Fig. \ref{fig_all}(h) is that the AVI-BoT model predicts higher possibility of outbreak at the shoreline than at the center of the water body. This is in conformity with \cite{locke198913} stating that Avian botulism is a water's edge disease, meaning that rarely are sick avi-fauna found away from the vegetation at the edge of the water or the original water’s edge.

For further analysis, we generated all year-round prediction maps using the proposed AVI-BoT model for 2019 shown in Table \ref{table_cs_klamath}. Some key observations from this case study are:
\begin{enumerate}
    \item The winter months (December-March) are marked by low prevalence of Avian botulism outbreaks. (Table \ref{table_cs_klamath}-(a), (b), (c), (l))
    \item As temperature begins to increase, the conditions turn conducive for the bacteria to develop, as evident from the increase in area with (probability $>$ 0.5) between April-June 2019. (Table \ref{table_cs_klamath} (d), (e), (f))
    \item During fall migration of birds (August-November), there exist news reporting \cite{klamath_2019} of avian botulism outbreak especially increasing in September 2019. The reason for this not showing up very effectively in the outbreak maps can be our reliance on NDWI maps since the water body could be having high levels of biomass which would result in shrinkage of 'water' area as per NDWI map and consequently the AVI-BoT prediction maps would not generate prediction on the masked area. (Table \ref{table_cs_klamath} (h)-(i)). 
    \item The news reporting of water release to Lower Klamath region in beginning of September 2019 \cite{water_release}, is confirmed by decreased avian botulism outbreak probability in the following months (Table \ref{table_cs_klamath} (j)-(k)). 
\end{enumerate}

\begin{table*}
\caption{Prediction maps for Lower Klamath National Wildlife Refuge, U.S.A. for 2019 using proposed AVI-BoT model.}
\label{table_cs_klamath}
\makebox[\linewidth]{
\begin{tabular}{|p{1.5in}|p{0.75in}|p{0.75in}|p{0.76in}|p{0.75in}|p{0.75in}|p{0.75in}|p{0.8in}}
\cline{1-7}
\textbf{AVI-BoT generated Prediction Map}  &  
(a) 13 Jan \includegraphics[height=0.6in]{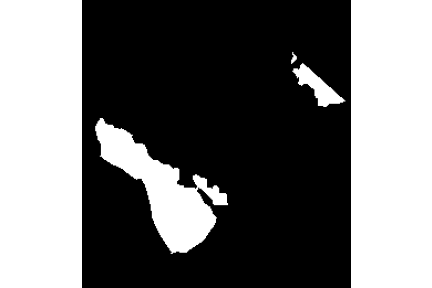} & 
(b) 19 Feb \includegraphics[height=0.6in]{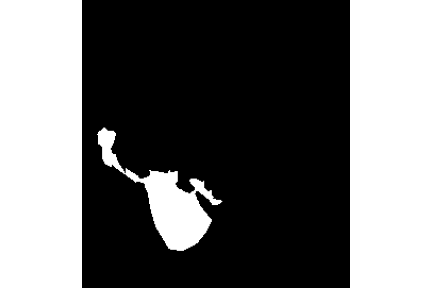}& 
(c) 14 Mar \includegraphics[height=0.6in]{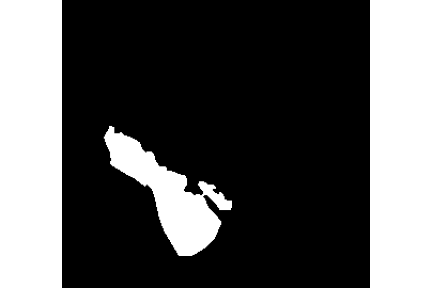} & 
(d) 18 Apr \includegraphics[height=0.6in]{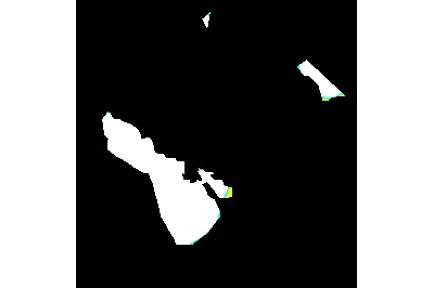}  & 
(e) 10 May \includegraphics[height=0.6in]{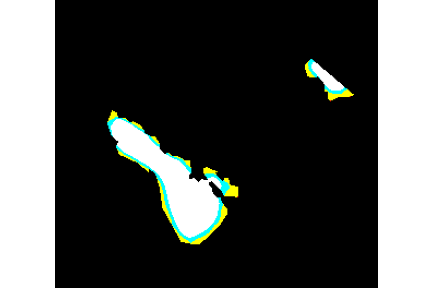}  & 
(f) 12 Jun \includegraphics[height=0.6in]{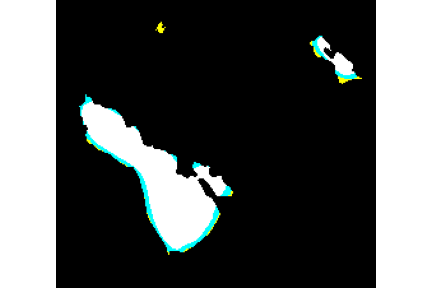} &
\multirow{4}{*}{\includegraphics[height=2.1in]{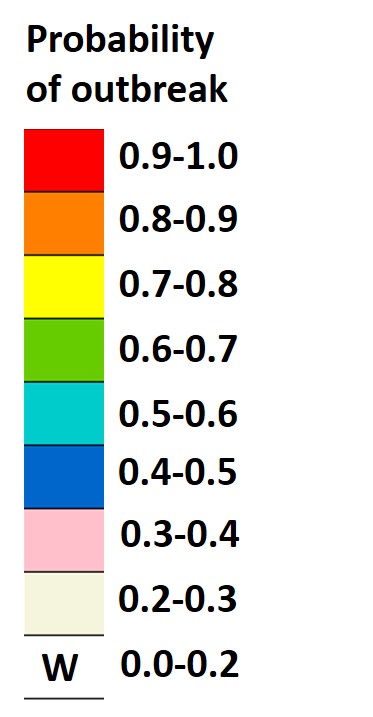}}\\
\cline{1-7}
\textbf{AVI-BoT Predicted Outbreak Probability}& Low & Low & Low & Low-Moderate & Moderate-High & Moderate & \\ 
\cline{1-7}
\textbf{Temp[Max,Min](C) Rainfall(mm)\cite{NOAA}} & \makecell{[5.9,-4.3] \\ 92.7} & \makecell{[1.0,-8.1] \\ 168.4} & \makecell{[7.6,-4.9] \\ 55.4} &  \makecell{[12.5,-0.1] \\ 95.0} &   \makecell{[16.8,2.1] \\ 57.4}&  \makecell{[22.6,4.4] \\ 7.4} \\
\cline{1-7}
\textbf{Possible hypothesis, qualitative analysis } & No outbreak in winter months & No outbreak in winter months & No outbreak even with onset of summer & Temp. begins to rise, conditions may become favourable & Temp. further increases and with less inflow of fresh water, conditions remain favourable; 
No cases reported & Conditions improve compared to May &  \\ 
\cline{1-7}
\\
\cline{1-7}
\textbf{AVI-BoT generated Prediction Map} &  
(g) 19 Jul \includegraphics[height=0.6in]{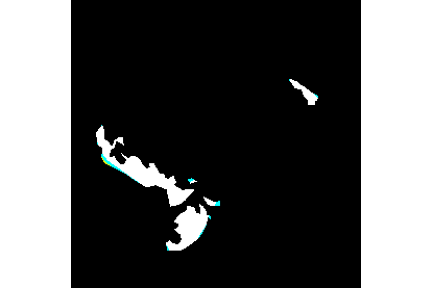}  & 
(h) 16 Aug \includegraphics[height=0.6in]{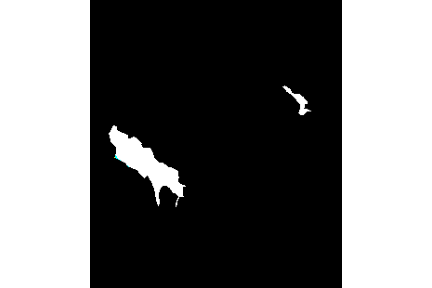} & 
(i) 12 Sep \includegraphics[height=0.6in]{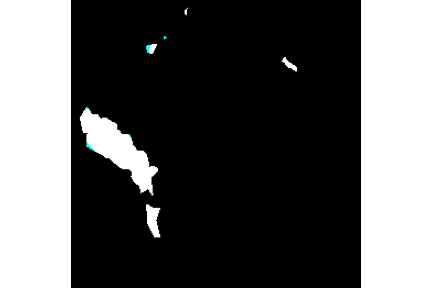} & 
(j) 22 Oct \includegraphics[height=0.6in]{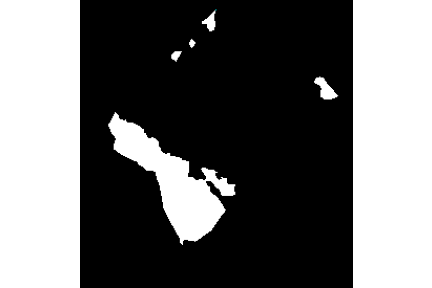}  & 
(k) 11 Nov \includegraphics[height=0.6in]{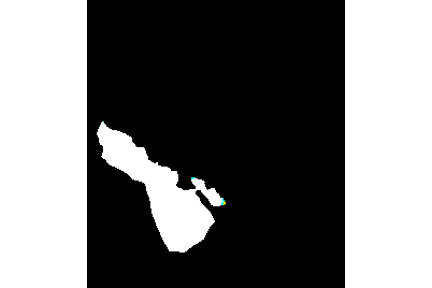}  & 
(l) 24 Dec \includegraphics[height=0.6in]{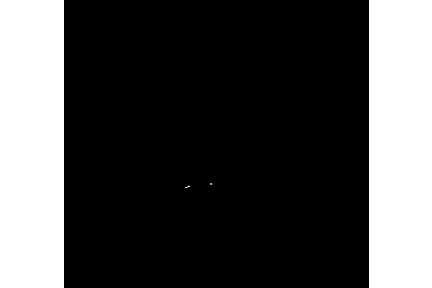} \\
\cline{1-7}
\textbf{AVI-BoT Predicted Outbreak Probability}& Low-Moderate & Low & Low-Moderate & Low & Low-Moderate & Low & \\ 
\cline{1-7}
\textbf{Temp[Max,Min](C) Rainfall(mm)\cite{NOAA}} & \makecell{[26.1,7.0] \\ 1.0} & \makecell{[27.3,8.2] \\ 18.5} & \makecell{[19.2,3.4] \\ 64.5} &  \makecell{[13.0,-2.8] \\ 22.9} &   \makecell{[11.1,-2.6] \\ 26.9}&  \makecell{[3.8,-4.5] \\ 82.8}& \\
\cline{1-7}
\textbf{Possible hypothesis, qualitative analysis } & Area with blue prediction (probability between 0.5 and 0.6) decrease & Increased rain may have eased conditions or presence of biomass could have shrunk NDWI,thus prediction may not be high & Presence of biomass could have shrunk NDWI map,thus prediction may not be high. Cases reported \cite{klamath_2019} & Water release \cite{water_release} may have led to low outbreak prediction & Water release \cite{water_release} may have led to low outbreak prediction & Negligible water mask created as per NDWI map &  \\ 
\cline{1-7}
\end{tabular}}
\end{table*}

Exhaustive temporal-analysis as shown can help monitor contributory factors round-the-year and timely issue a pre-emptive warning, thereby helping authorities to act and save migratory birds in large numbers.

\begin{figure}[H]
\makebox[1.1\linewidth]{
\includegraphics[width=7in]{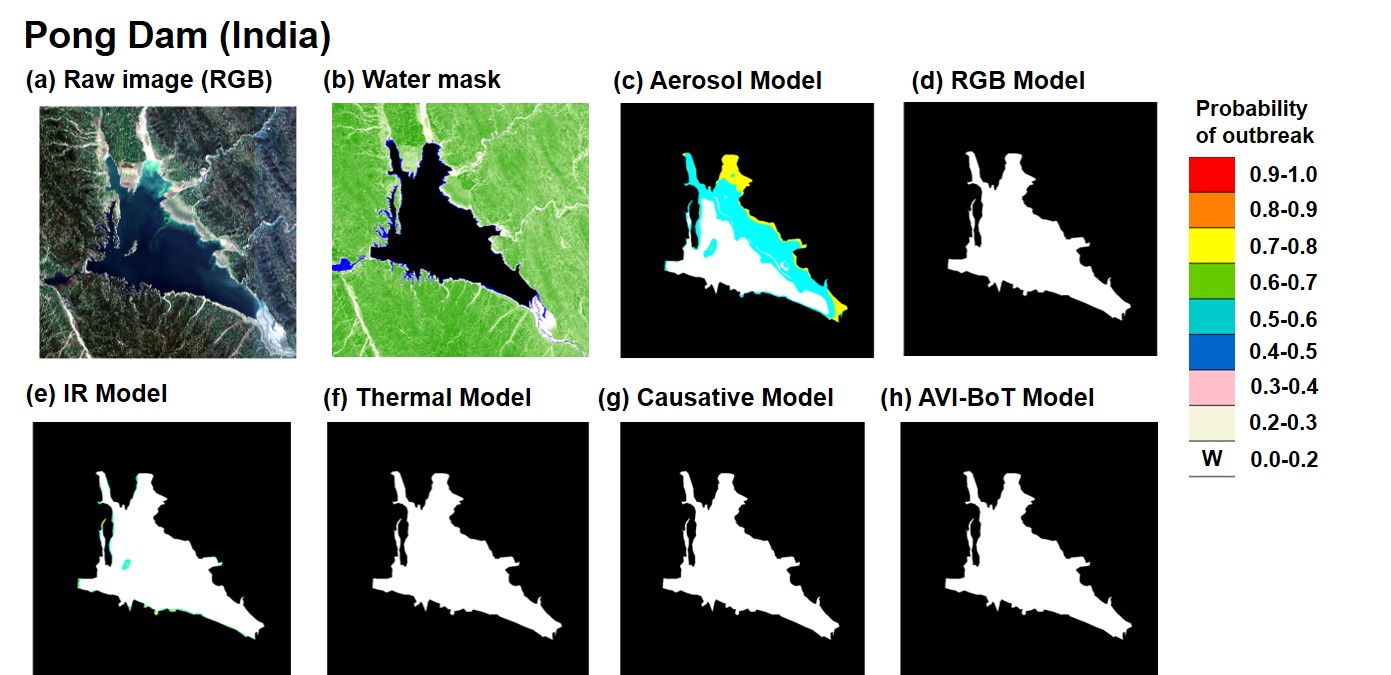}}
\caption{Prediction results on Pong Dam, India for 27th January 2021 ((a)-(h)).}
\label{fig_all_pong}
\end{figure}

\subsubsection{Test Case Study II - Pong Dam, India}
Pong Dam (Himachal Pradesh, India) saw massive bird deaths in January 2021 with a count of nearly 5000 per month which was contained in early February 2021 \cite{pong_news}. The reason for this massive die-off was earlier speculated to be Avian Botulism but was later confirmed to be Avian Flu. For this case study, this sample is hence treated as 'negative'. 

Fig. \ref{fig_all_pong} (b) shows the water mask used for prediction in confirmation with NDWI mask. Aerosol model (Fig. \ref{fig_all_pong} (c)) and IR model (Fig. \ref{fig_all_pong} (e)) yield a positive prediction while all other models show a negative prediction. (Fig. \ref{fig_all_pong} (d), (f), (g), (h)).

\subsubsection{Test Case Study III - Langvlei and Rondevlei lakes, South Africa}
Langvlei and Rondevlei lakes (South Africa) are two wilderness lakes in vicinity of each other documented to show Avian Botulism related deaths since 2015 \cite{govender2019outbreak}. For this case-study, we rely on \cite{govender2019outbreak} to provide us with actual on-ground lake-wise and month-wise mortality as well as information about physiological parameters. Due to limitation over data overlapping with the paper \cite{govender2019outbreak}, we use data samples over three months in 2017 - January, February and May. 

\begin{figure}[H]
\makebox[1.1\linewidth]{
\includegraphics[height=3.2in,width=7in]{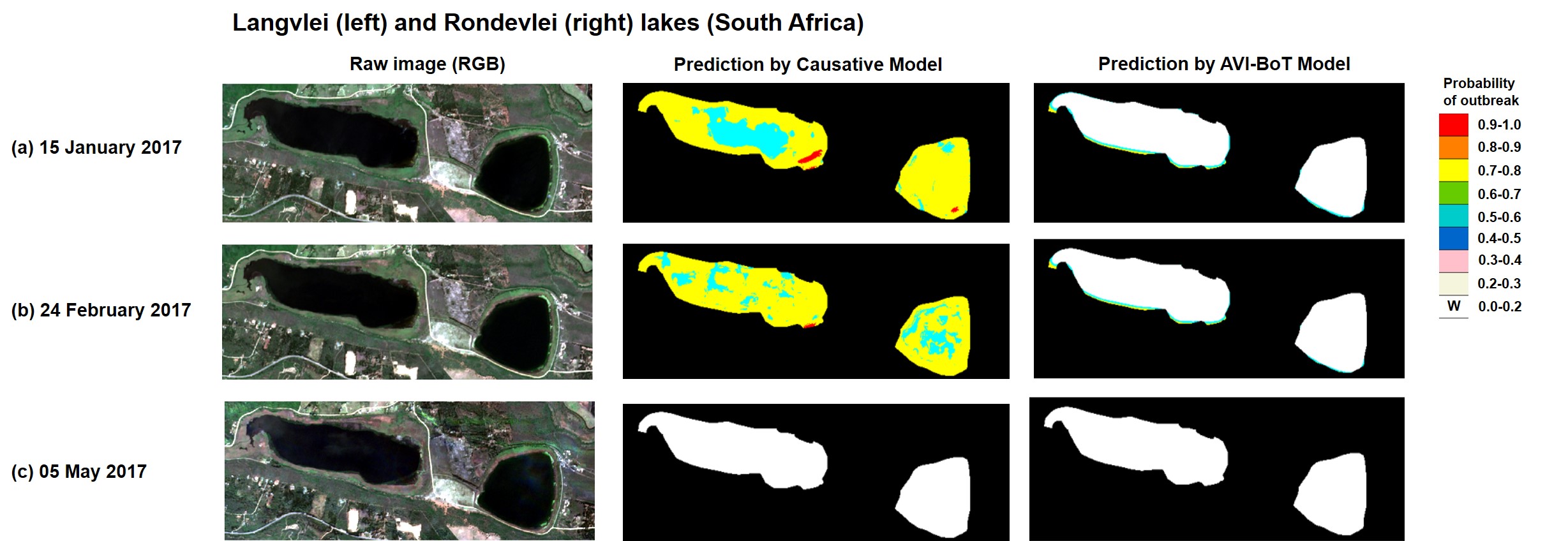}}
\caption{Prediction results on Langvlei and Rondevlei lakes, South Africa for (a) 15 January 2017, (b) 24 February 2017 and (c) 05 May 2017 - for both Causative Factors based Model and AVI-BoT Model.}
\label{fig_all_sa}
\end{figure}

The trends deduced from the paper \cite{govender2019outbreak} discussed below are confirmed with predictions from the trained models (Fig. \ref{fig_all_sa}) as:

\begin{enumerate}
    \item The paper documents a mortality only for Langvlei lake in the first five months of 2017 and zero mortality for Rondevlei lake. This is confirmed by the AVI-BoT model predictions as the outbreak probability for Langvlei lake by AVI-BoT model is higher than Rondevlei lake (Fig. \ref{fig_all_sa} (a),(b)). It may be noted that the prediction by AVI-BoT over Langvlei lake is itself not a very highly positive prediction which may be attributed to the low mortality (less than 50) over Langvlei lake in 2017 compared to other years.
    \item The paper documents a 0 mortality for Rondevlei lake over the first 5 months of 2017, which is nearly confirmed as AVI-BoT model predicts a very slight region of (probability $\sim$ 0.5-0.6) over the edge in Fig. \ref{fig_all_sa} (a), (b) and negative prediction over Fig. \ref{fig_all_sa} (c).
    \item The paper documents decreasing mortality from January to May 2017 which is also evident from the decreasing area of high probability of outbreak (probability $>$ 0.5) from Fig. \ref{fig_all_sa} (a) - (c). 
    \item The paper documents physiological parameters like Temperature, Salinity, Dissolved Oxygen for the two lakes across the time period. Since these parameters themselves have been used to build the Causative Factors model, the predictions by Causative Factors model for the 3 data samples are compared with AVI-BoT Model. For the period of January-May 2017, the temperature increases, salinity decreases, and dissolved oxygen increases. While these physiological parameters vehemently support the on-ground trend of mortality reduction, the Causative Factors based Model is completely inaccurate over Rondevlei lake showing a positive prediction despite having a negative label, and gives an all-encompassing positive prediction over Langvlei lake which increases (area with probability between 0.7 and 0.8) between Fig. \ref{fig_all_sa} (a) to (b), which is contrary to the on-ground mortality reduction which is documented.
    \item The paper documents a very low mortality in May 2017, which can also be seen as the Fig. \ref{fig_all_sa} (c) shows a negative prediction by both Causative Factors model and AVI-BoT model.
\end{enumerate}

\section{Analysis} \label{analysis}
The proposed AVI-BoT model outperforms the proposed literature-inspired Causative Factors model with a significantly higher accuracy. It uses Aerosol band to capture the constituents of shallow water, Visible band is the abstraction of salinity and organic matter, Infra-Red bands capture other crucial factors like pH, and Thermal bands are a direct input for temperature. Thereby, covering all relevant physical, chemical and biological parameters/features through learning fused multi-spectral satellite data.

\begin{figure}[H]

\makebox[1.1\linewidth]{
\includegraphics[width=0.8\linewidth]{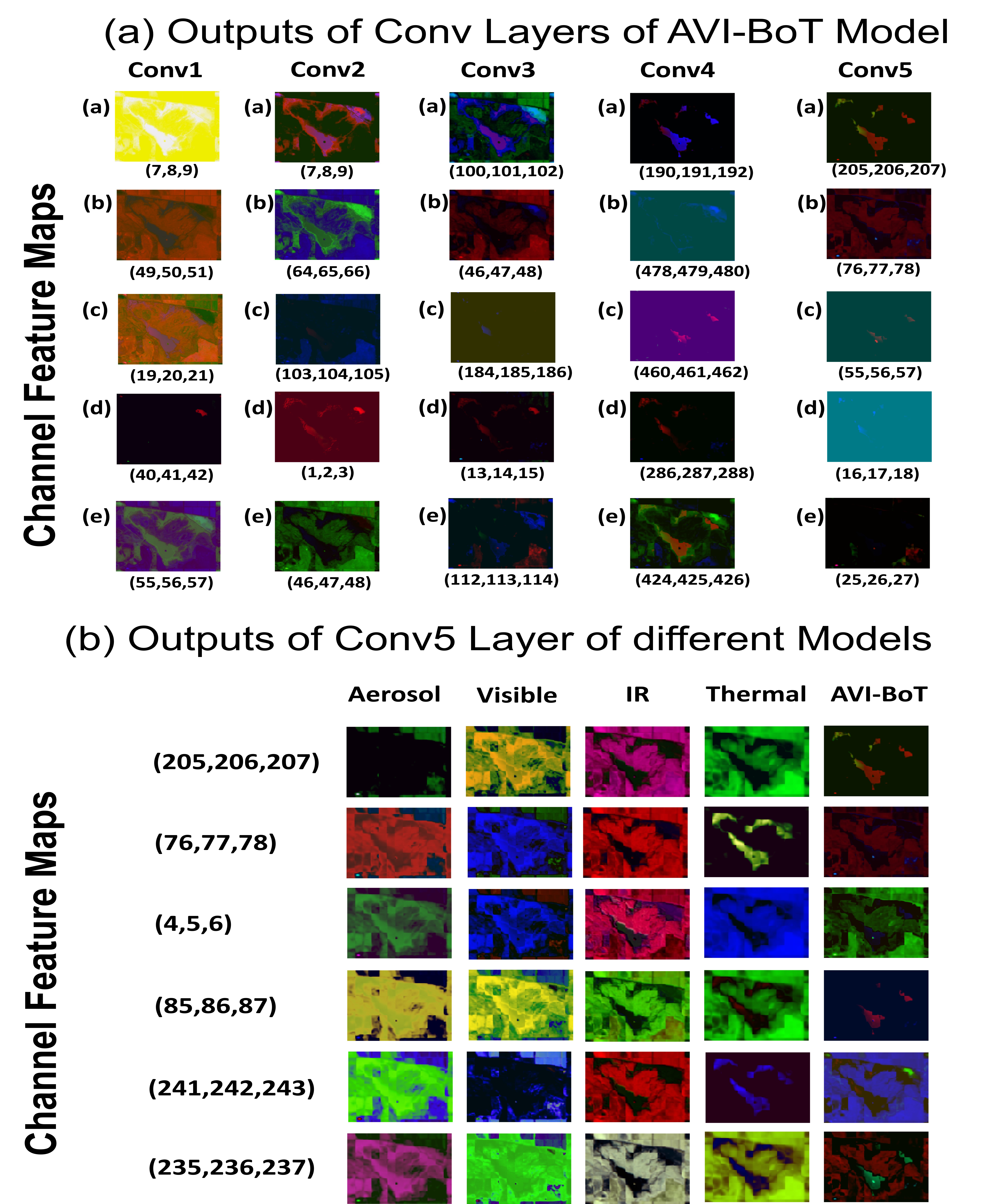}}
\caption{Illustration of progress of the feature-space learning through the convolutional layers,  by proposed CNN models, on Lower Klamath National Wildlife Refuge (U.S.A.) sample input data (i.e. 10-band fused image dated 9th October 2020). Since convolutional layers have multiple channels, three successive channels are stacked in each row (viewed in false RGB) as a visual aid for human interpretation. 
(a) Conv. layers 1-5 using AVI-BoT model. The (x,y,z) below each image represent stacking of feature maps of xth, yth, zth layer (in false RGB). (b) Conv. layer 5 using five different CNN models.}
\label{fmaps_avibot}
\end{figure}

To better understand AVI-BoT's learning of feature-space representations, we examine the progression of features learnt by AVI-BoT through successive convolution layers. These visualizations (see Fig. \ref{fmaps_avibot} (a)) have been generated using Lower Klamath National Wildlife Refuge (U.S.A.) as sample inference data point for 9th October 2020. Analysis shown in Fig. \ref{fmaps_avibot} (a) helps us to understand the gradual progression of learning the feature space through Conv1 to Conv5 layers. To understand the learning at each convolutional layer, we have hand-picked the most useful feature representations and their representations are explained below (see Fig. \ref{fmaps_avibot} (a)).

\begin{enumerate}
    \item Fig. \ref{fmaps_avibot} (a) Conv1 Layer - learns basic object separation. (a), (b) shows that it is responsible for learning basic background and foreground. (c) It further learns how to distinguish between various objects on a scene, (d) some feature maps get activated on solitary objects, (e) it begins to distinguish between objects which belong to the same class (eg. water).
    
    \item Fig. \ref{fmaps_avibot} (a) Conv2 Layer - learns more defined tasks like edge detection. Both (a), (b) show that the layer begins to learn edge detection as can be seen over the edges of the lake. While (d) shows strong activation by the water area in a single feature layer, (c), (e) show that the layer begins to learn gradation within the water area in the lake.
    
    \item Fig. \ref{fmaps_avibot} (a) Conv3 Layer - begins to apply the edge detection and gradation within the water body area. Both (a), (b) show defined response by different feature layers over edges of the lake, and various water areas. Both (c) and (d) show how the network learns to fine-tune features when going deeper by narrowing down the area by which it is activated. (e) gives an idea of how the network differentiates between the characteristics shown by various regions of the water body.
    
    \item Fig. \ref{fmaps_avibot} (a) Conv4 Layer - specializes on the gradation and picks up various areas within the water body with defining unique characteristics. (a) shows a certain water area with a very defined  characteristic which the network learns to pick up which is even seen in Conv5 Layer (a) - this major unique characteristic picked up by AVI-BoT Conv4 and Conv5 Layer translates to the final prediction map output and can be said to be a very defining feature map layer for AVI-BoT Model. (b), (c), (d) show the various water body areas that the AVI-BoT model sees as distinct from each other while looking similar in many other dimensions like (Aerosol, Visible, Infra-Red, Temperature) as will be seen below. (e) shows a feature map with many such areas superimposed. 
    
    \item Fig. \ref{fmaps_avibot} (a) Conv5 Layer - fine-tunes over Conv4 layer output to give very defined and precise activations by the various regions within the water body as can be seen from (a) - (e). This unique feature space seen by AVI-BoT is a crucial determinant of the final prediction map.
    
\end{enumerate}

To further examine the learnt feature space, we delve into a deeper feature analysis and additionally train four direct spectral models ((i) Aerosol, (ii) Visible (3-bands), (iii) IR (3-bands), (iv) Thermal (3-bands)). We compare the features learnt by each of the direct spectral models and try to interpret/hypothesize their contribution in the learnt feature space of AVI-BoT. The following observations can be drawn from Fig. \ref{fmaps_avibot} (b):

\begin{enumerate}
    \item Aerosol model's feature maps have almost uniform activations over the entire surface of the lake. This can be a reason of poor performance when compared to other models despite capturing water constituents well.
    \item Visible model's feature maps too have almost uniform activations which can be a reason for the incorrect predictions for Lower Klamath National Wildlife Refuge. 
    \item Thermal model activations have more gradation as compared to Aerosol and Visible model which is also evident from Thermal model atleast giving a correct prediction in two test case-studies (Lower Klamath National Wildlife Refuge and Pong Dam). This is despite having low spatial resolution (100 m/1000 m) as well as near-uniform temperature in areas in vicinity. 
    \item IR model have high gradation activation maps with activation on the shoreline as can be seen from (4,5,6), (76,77,78), (85,86,87). This can attributed to the IR bands capturing pH levels, vegetation and organic matter very well.
    \item The AVI-BoT model has a strong spatial activation map where various regions activate distinct feature channels as can be seen from the AVI-BoT Column where each feature map corresponds to a different region of the water body.
    \item The AVI-BoT model feature maps, to some extent, can be seen as made up of contributions from each of the (Aerosol, Visible, Infra-Red, Thermal) feature maps. Eg. the activation observed at the shore of the lake for AVI-BoT model is dependent majorly on the IR model.
    \item The lower right portion of lake in AVI-BoT model's feature map can be seen as determined from both Thermal and Visible feature maps (4,5,6).
    \item Temperature seems to be have dominant role in the shallow region like upper region of the bigger water body (eg. Thermal feature map of (85,86,87) contributes to the AVI-BoT model feature map of (241,242,243)).
    \item From all the feature maps, it is evident that AVI-BoT model captures features of the water body in a unique hyperplane which leads to differentiation of water body and capturing of areas which are not traditionally distinguished in the (A,V,I,T) feature maps. Each feature map of AVI-BoT can be seen to capture a unique area within the water body. And areas such as marked in green in (205,206,207) are instrumental in determining outbreak probability as this is the main area in Fig. \ref{fig_all} (h) predicted to have high outbreak probability. 
\end{enumerate}

\begin{figure}[H]
\centering
\makebox[1.05\linewidth]{
\includegraphics[height=3.45in]{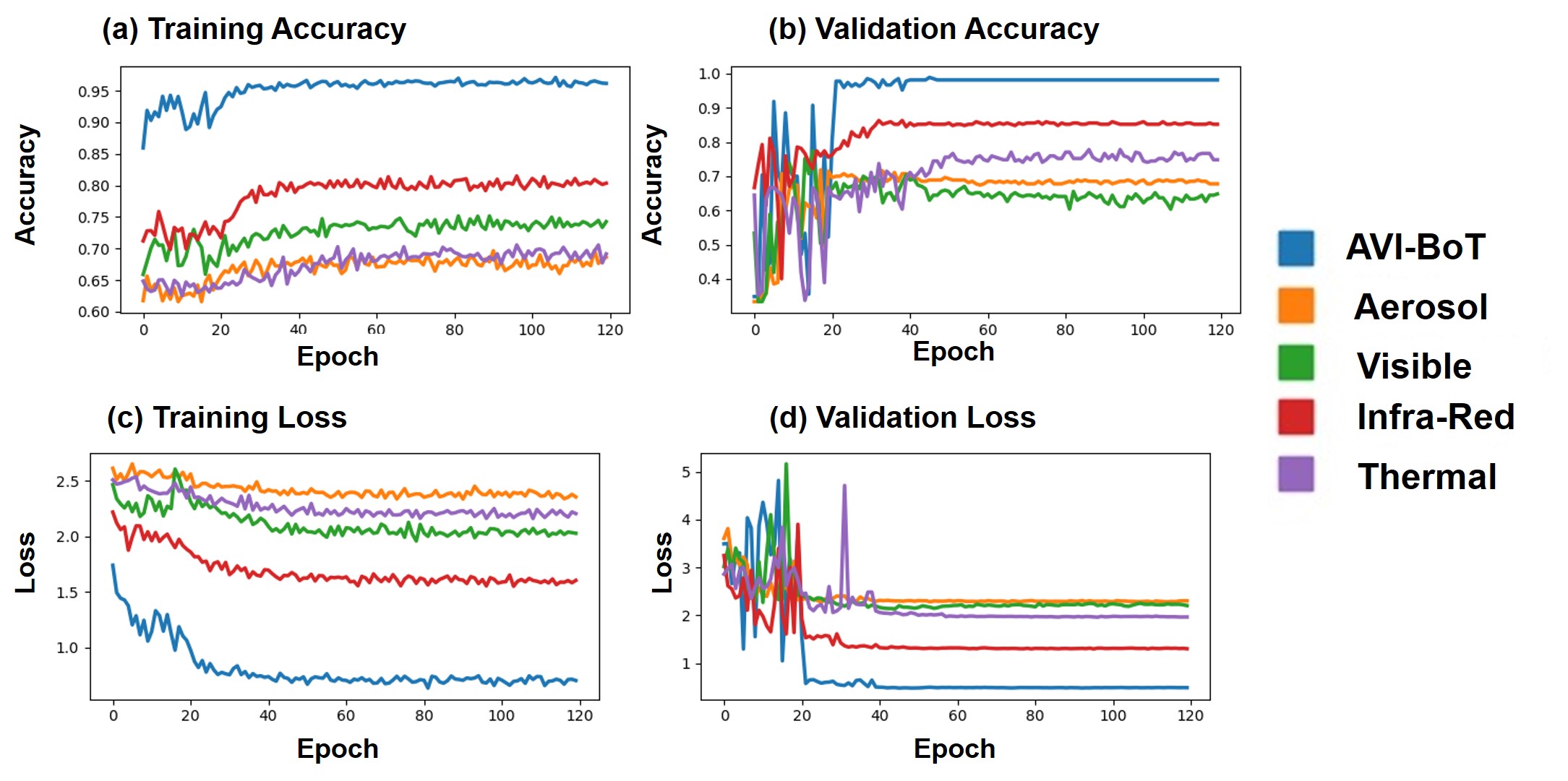}}
\caption{Epoch-wise scores for various spectral feature input representations: (a) Training Accuracy, (b) Validation Accuracy, (c) Training Loss, and (d) Validation Loss.}
\label{fig_train_analysis}
\end{figure}

Fig. \ref{fig_train_analysis} shows the accuracy and loss plots for training and validation for the four direct models compared with AVI-BoT model. All models lag far behind the AVI-BoT model, with IR model being the second best in performance. This is also validated by the IR column in Fig. \ref{fmaps_avibot} (b) contributing nearly the highest influence in activating the network to yield a prediction.

From the test-case studies - Lower Klamath National Wildlife Refuge (Fig. \ref{fig_all}) and Pong Dam (Fig. \ref{fig_all_pong}), it is evident that Aerosol model and RGB Model give inaccurate predictions which can be expected as it can be the result of a biased prediction relying solely on the water texture (aerosol) or optical features (visible). Thermal Model, despite accurately giving a positive/negative result shows it's high susceptibility to slight variations, and as it is just one of the factors leading to Avian botulism, hence a model solely reliant on temperature will not be the optimal solution. The IR model too is not accurate despite high influence on feature maps and being second best in performance after AVI-BoT because even though it yields a correct prediction for Lower Klamath National Wildlife Refuge with a moderate prediction over the shoreline, it gives an inaccurate result for Pong Dam. 

A correct comparison of Causative Factors based model and AVI-BoT model can be made only after accounting for the third case study - Langvlei and Rondevlei lakes (Fig. \ref{fig_all_sa}) in addition to the first two case studies. While Causative Factors based model can be considered to declare true negatives in very clear situations (such as Pong Dam where avian botulism was not at all present, or 05 May 2017 sample of Langvlei and Rondevlei lake), it suffers on two counts - failing to accurately predict for a complex case like Rondevlei lake (January and February 2017 samples) and giving an overwhelmingly positive blanket prediction over the entire lake. On the other hand, all three case-studies show that AVI-BoT not only gives an accurate prediction of positive or negative but also preserves spatial information which is useful for conservation and prevention efforts. Future studies which document on-ground physiological parameters with actual clinical diagnosis can help fine-tune the prediction of AVI-BoT for getting higher spatial accuracy. Further, reliable water segmentation techniques for complex cases such as presence of biomass in water and standardized definition such as NDWI or moisture index can be used to generate training samples which can help remove any land areas that may creep in otherwise due to being close to shoreline or the water body being very small in size.

\section{Conclusion}
This interdisciplinary study presents the intersection of - wildlife conservation, satellite based remote-sensing technology and advanced computational techniques driven by AI and deep learning. We build a \textit{Bird-Area Water Bodies} Dataset (BAWD) of multi-spectral satellite images of bird-habitats covering 16 topographically diverse global locations over the period 2016-2021, with ground-truth labels backed by on-ground reportings confirming the outbreak occurrence. We further present AVI-BoT, an intelligent AI model with validation accuracy of 0.989, to forecast Avian botulism outbreaks. We also validate our technique on three test case study locations. The proposed model generates spatial decision output map that is closely aligned with actual on-ground clinical diagnosis based reports. We also perform an ablation study to examine the evolution of the feature space. Proposed predictive methodology provides an efficient and scalable first-of-its-kind monitoring technique which can be used to pre-emptively protect the migratory fauna and interdependent food webs.
\bibliographystyle{naturemag}
\bibliography{ref}

\end{document}